\documentstyle[epsf,twocolumn,aps,prb]{revtex} 

\def\prl#1#2#3{Phys. Rev. Lett. {\bf #1}, #2 (#3)} 
\def\prb#1#2#3{Phys. Rev. B {\bf #1}, #2 (#3)} 
\def\ibid#1#2#3{{\it ibid}. {\bf #1}, #2 (#3)} 
\def\pc#1#2#3{Physica (Amsterdam) {\bf #1C}, #2 (#3)}

\begin{document}

\draft 

\twocolumn[\hsize\textwidth\columnwidth\hsize\csname 
@twocolumnfalse\endcsname 

\title{
Bulk magnetic properties and phase diagram of Li-doped La$_2$CuO$_4$: \\
Common magnetic response of hole-doped CuO$_2$ planes 
}

\author{T. Sasagawa,$^{1,2}$ P. K. Mang,$^{2}$ O. P. Vajk,$^{3}$ 
 A. Kapitulnik,$^{2,3}$ and M. Greven$^{2,4}$}

\address{$^{\rm 1}$Department of Advanced Materials Science, 
 University of Tokyo, 7-3-1 Hongo, Bunkyo-ku, Tokyo 113-0033, Japan} 
\address{$^{\rm 2}$Department of Applied Physics, 
 Stanford University, Stanford, California 94305} 
\address{$^{\rm 3}$Physics, 
 Stanford University, Stanford, California 94305} 
\address{$^{\rm 4}$Stanford Synchrotron Radiation Laboratory, 
 Stanford University, Stanford, California 94305} 

\date{Published: Physical Review B {\bf 66}, 184512 (2002)}
\maketitle

\begin{abstract}
Although La$_2$Cu$_{1-x}$Li$_x$O$_4$ (Li-LCO) differs from 
La$_{2-x}$Sr$_x$CuO$_4$ (Sr-LCO) in many ways 
(e.g., the absence of metallic transport, high-$T_c$ superconductivity, 
and incommensurate antiferromagnetic correlations), 
it has been known that certain magnetic 
properties are remarkably similar. 
The present work establishes the detailed bulk magnetic phase diagram 
of Li-LCO ($0\le x \le 0.07$), 
which is found to be nearly identical to that of lightly-doped Sr-LCO, 
and therefore extends the universality of the phase diagram to 
hole-doped but nonsuperconducting cuprates. 
\end{abstract}

\vspace*{5mm}
] 
\narrowtext 


\section{INTRODUCTION} 
Depending on the nature and concentration of dopants, La$_2$CuO$_4$ (LCO) 
displays a wide variety of phenomena such as antiferromagnetism, 
spin glass (SG) behavior, an anomalous metallic response, and 
high-temperature superconductivity (HTS). 
Undoped LCO contains weakly coupled CuO$_2$ 
planes and exhibits antiferromagnetic (AF) order of the Cu$^{2+}$ 
spin-1/2 moments below $T_N \approx 325$ K. 
Replacement of Cu$^{2+}$ by nonmagnetic Zn$^{2+}$ or Mg$^{2+}$ models 
random spin dilution, leading the system into a disordered state 
at doping concentrations above $\sim\,40\%$.\cite{Owen} 
On the other hand, substitution of divalent alkaline earth cations for 
La$^{3+}$ or the introduction of excess interstitial oxygen 
introduces hole charge carriers into the CuO$_2$ planes 
which frustrate the spin system.\cite{L60p1330} 
A concentration of $x=0.02$ in La$_{2-x}$Sr$_x$CuO$_4$ (Sr-LCO) 
is enough to destroy the AF order; this value is one order of 
magnitude smaller than for the spin-dilution case. 
Spin freezing has been found at low temperatures 
in the AF doping regime ($x < 0.02$),\cite{L71p2323} with recent 
direct evidence for electronic phase separation.\cite{Matsuda} 
Further doping ($x > 0.02$) leads to the emergence of 
a SG phase,\cite{B46p3179,L75p2204,B62p3547} 
followed by superconductivity for $x \approx 0.06 - 0.25$. 
SG order is found to coexist with superconductivity,\cite{SG+SC} 
with no apparent anomaly in the SG temperature at the doping 
level $x \sim0.06$ at which superconductivity first 
occurs.\cite{SG+SC,L80p3843,PC341p843} 
The phase diagram of the substitutionally-doped bilayer 
material Y$_{1-x}$Ca$_x$Ba$_2$Cu$_3$O$_6$ (Ca-YBCO) 
closely resembles that of Sr-LCO,\cite{L80p3843} 
and spin freezing has also been found in 
LCO (Ref. 11) and YBCO (Ref. 12) 
doped with excess oxygen.

The extrapolated disappearance of SG order in Sr-LCO and Ca-YBCO 
appears to coincide with the doping level 
at which the normal state pseudogap extrapolates to zero, 
and it has been suggested 
that this might be consistent with predictions involving 
quantum criticality.\cite{PC341p843,Chakravarty,Andergassen} 
In the $d$-density wave picture of HTS,\cite{Chakravarty} 
the reason for the lack of a genuine phase transition 
at the pseudogap temperature is that the disorder present in all 
existing cuprates corrupts the $d$-density wave order and transforms 
the transition into the low-temperature SG transition. 
While the freezing of $d$-density wave fluctuations is one proposal 
for the origin of the SG in doped cuprates, there also exist other 
interpretations.\cite{B46p3179,L75p2204,B62p3547,Kivelson,Gooding,L85p836,EPL56p870} 
For example, it has been argued that the glassiness found in doped 
Mott insulators may be self-generated, due to 
the competition between interactions on different length scales, 
and that quenched disorder may merely 
further stabilize SG order.\cite{L85p836} 
The most discussed scenario has been the 
so-called cluster SG,\cite{B46p3179,L75p2204} with holes on the cluster 
boundaries and in the clusters giving rise to 
SG physics and to the experimentally observed 
incommensurate spin correlations.\cite{B62p3547} 

Given the enormous interest in the connection between 
magnetic correlations and HTS, it would be valuable to investigate the 
detailed magnetic properties of related, nonmetallic 
materials such as La$_2$Cu$_{1-x}$Li$_x$O$_4$ 
(Li-LCO).\cite{B37p111,B54p9538,B54p12014,L81p2791,Bao} 
Since Li$^+$ not only provides one hole carrier, but also removes 
a Cu$^{2+}$ spin, this system experiences the dual effects of spin 
dilution and frustration. 
Unlike Sr-LCO, Li-LCO does not superconduct\cite{B37p111} 
and shows no evidence of incommensurate AF correlations.\cite{Bao} 
However, early work on polycrystalline samples 
reported a rapid suppression of AF order with 
doping,\cite{B54p9538,B54p12014,L81p2791} 
similar to Sr doping rather than Zn doping, and evidence for spin freezing 
in the N\'eel state.\cite{L81p2791} 
The remarkable similarity of magnetic properties of lightly doped Li-LCO 
with those of Sr-LCO has been interpreted as due to new 
collective behavior of the holes.\cite{L81p2791} 
Therefore, in connection with HTS, 
the detailed magnetic properties and phase diagram 
of Li-LCO are of considerable interest. 
Experimentally, spin freezing 
is observed at a different temperature, depending on the time scale 
of the probe.\cite{B62p3547,Markiewicz} 
Unlike NQR,\cite{L71p2323,L81p2791} $\mu$SR,\cite{L80p3843,PC341p843} 
and neutron scattering,\cite{Matsuda,B62p3547} 
magnetometry using a superconducing quantum interference device (SQUID) 
is essentially a static probe, allowing a more accurate extraction 
of the SG transition temperature $T_{\rm sg}$.\cite{L75p2204,B62p3547} 
Our results for lightly-doped samples demonstrate 
the feasibility of using bulk magnetometry to extract 
spin freezing temperatures in the N\'eel regime. 
The present magnetometry study establishes the existence 
of a nearly quantitative agreement of the bulk magnetic phase diagram 
of Li-LCO with that of Sr-LCO. 
Since the strength and type of the disorder as well as the charge 
transport differ significantly in these two materials, 
this finding places constraints on the origin of the SG degrees 
of freedom in hole-doped cuprates. 

\section{EXPERIMENTAL} 
Using the traveling-solvent floating-zone method,\cite{L80p4297} 
we have succeeded 
in growing large single crystals of La$_2$Cu$_{1-x}$Li$_x$O$_4$ 
(0 $\leq $ $x$ $\leq $ 0.07). 
The crystal axes were precisely determined by the x-ray Laue 
backscattering technique. 
In order to eliminate possible hole doping by excess oxygen, 
the crystals were carefully heat treated under reducing conditions. 
The Li concentrations were estimated to within $\pm 0.003$ from 
x-ray diffraction measurements of the lattice constants.\cite{B54p12014} 
For a few samples, we confirmed this estimate using neutron diffraction 
to determine structural and N\'eel 
transition temperatures.\cite{B54p12014} 
All magnetometry data reported here were taken with a commercial SQUID 
magnetometer with the magnetic field along the tetragonal $a$-axis. 

\begin{figure}[b]
 \vspace*{8mm}
 \centerline{\epsfxsize=2.5in\epsfbox{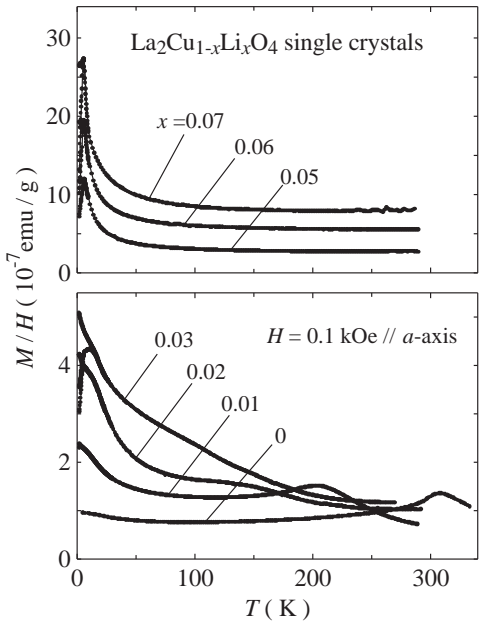}}
 \vspace*{5mm}
 \caption{Magnetic susceptibility of La$_2$Cu$_{1-x}$Li$_x$O$_4$ 
 with a systematic change of $x$. 
 A magnetic field $H = 0.1$ kOe was applied parallel to the tetragonal 
 $a$-axis.}
\label{mags}
\end{figure}

\section{RESULTS AND DISCUSSIONS} 
Figure~\ref{mags} summarizes the temperature and doping dependence 
of the magnetic susceptibility of La$_2$Cu$_{1-x}$Li$_x$O$_4$, 
taken with a field of $H$ = 0.1 kOe along the tetragonal $a$-axis. 
The data exhibit several features which systematically shift 
with doping up to $x$ = 0.03, and change their nature 
above this doping level. 
Earlier powder studies had demonstrated a rapid suppression of
the AF order with Li doping,\cite{B54p9538,B54p12014,L81p2791} 
and the Li concentration at which N\'eel order is lost 
corresponds to $x \sim$ 0.03. 
The doping dependence of the N\'eel temperature determined 
by local\cite{B54p9538,L81p2791} and bulk\cite{B54p12014} 
magnetic probes agree quite well, 
and until now this has been the only magnetic phase boundary known 
in the Li system. 
However, the present results indicate that the phase diagram in Li-LCO 
is more complicated than what has been determined from powder samples. 

\begin{figure}[b]
 \vspace*{4mm}
 \centerline{\epsfxsize=2.8in\epsfbox{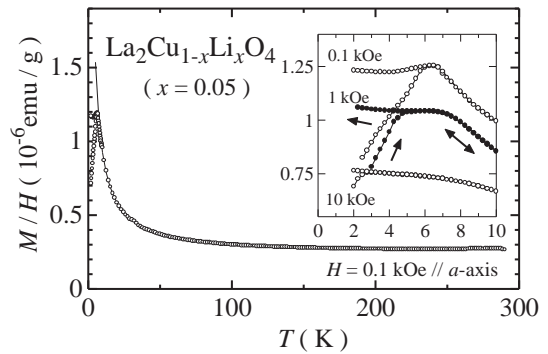}}
 \vspace*{0.3cm}
 \caption{Magnetic susceptibility of La$_2$Cu$_{0.95}$Li$_{0.05}$O$_4$. 
 A magnetic field $H = 0.1$ kOe was applied parallel to the tetragonal 
 $a$-axis. 
 The solid curve is a Curie-Weiss fit of the high-temperature data. 
 The inset shows the low-temperature behavior at various fields. 
 Arrows indicate the $T$-scan directions.}
\label{mag05}
\end{figure}

$\mu$SR results indicate magnetic order at low temperatures 
even up to $x\sim $ 0.10,\cite{B54p9538} but bulk magnetization 
measurements in powder samples only show spin paramagnetism 
for $x > 0.03$.\cite{B54p12014} 
The temperature dependent susceptibility for $x > 0.03$ 
can therefore be expressed by the extended Curie-Weiss formula: 
\begin{equation}
 \chi (T) = \chi _0 + \frac{C}{(T - {\it \Theta})},
 \label{suscept}
\end{equation}
where $\chi _0$, $C$, and $\it \Theta $ are the $T$-independent 
susceptibility, Curie constant, and Curie-Weiss temperature, respectively. 
Figure~\ref{mag05} shows $\chi (T)$ for $x = 0.05$. 
As in the earlier powder report,\cite{B54p12014} paramagnetism 
is observed in the high temperature regime. 
A fit to Eq.~(\ref{suscept}) (solid line in Fig.~\ref{mag05}) 
resulted in $\chi _0$ = 2.35 $\times $ 10$^{-7}$ emu/g, 
$C$ = 7.18 $\times $ 10$^{-6}$ emu$\cdot $K/g, and ${\it \Theta} <$ 0.5 K. 
We note that these parameters are roughly comparable with those Sr-LCO: 
$\chi _0$ = 0.4 $\times $ 10$^{-7}$ emu/g for $x$ = 0.04,\cite{L75p2204} 
$C$ = 2$-$5 $\times $ 10$^{-6}$ emu K/g, and $\it \Theta $ = 0 K 
for $x$ = 0.03$-$0.05.\cite{L75p2204,B62p3547} 

At low temperatures, $\chi (T)$ deviates significantly from a 
Curie-Weiss law and shows signatures of a SG. 
As described in detail below by the appropriate limits 
of the scaling function, 
the SG order parameter in the zero-field limit increases 
from zero upon cooling below $T_{\rm sg}$, 
which lead to a decrease of $\chi (T)$ for $T < T_{\rm sg}$. 
Furthermore, the resulting peak becomes broader as the applied magnetic 
field is increased due to the enhancement 
of the order parameter both above and below $T_{\rm sg}$. 
This behavior is demonstrated in the inset of Fig.~\ref{mag05}. 
From the peak in the lowest field ($H=0.1$ kOe) 
we determined $T_{\rm sg} = 6.2(1)$ K, which should be compared 
to $T_{\rm sg} = 5.0(5)$ K for Sr-LCO with $x=0.05$.\cite{B62p3547} 
Hysteresis below $T_{\rm sg}$, observed between zero-field-cooled and 
field-cooled curves, is indicated with the help of arrows 
in the inset of Fig.~\ref{mag05}. 
Such a behavior is also characteristic of a SG. 
Observing these features in either small crystals or polycrystalline 
samples would be difficult because they become obscured 
as the field is increased, or when the field 
direction is canted away from the CuO$_2$ plane. 

\begin{figure}[t]
 \centerline{\epsfxsize=2.8in\epsfbox{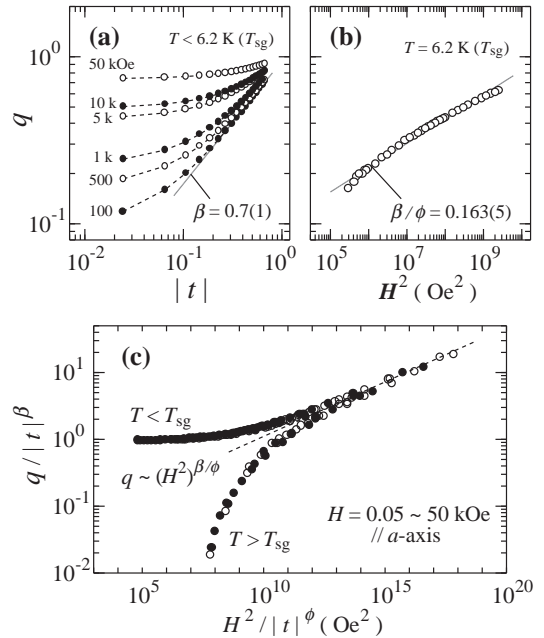}}
 \vspace*{0.3cm} 
 \caption{Scaling analysis of the data in Fig. 1. 
 SG order parameter $q$ as a function of 
 (a) reduced temperature $t$ and (b) field squared $H^2$. 
 (c) Scaling plot of the SG order parameter.}
 \label{scales}
\end{figure}

The large oriented single crystals used in the present study have 
enabled us to further characterize the SG state by means of 
a critical scaling analysis of the SG order parameter $q$, 
which is experimentally associated with the deviation of the observed 
equilibrium (field-cooled) susceptibility $\chi(T,H)$ 
from Curie behavior.\cite{L51p1704} 
Normalization to satisfy 0 $\leq q \leq$ 1 yields 
\begin{equation}
 q(T,H) = \left[\left(\chi_0+\frac{C}{T}\right) - \chi (T,H)\right]
 /\left(\frac{C}{T}\right).
 \label{order-para}
\end{equation}
Theoretically, the SG transition should obey 
a scaling relation\cite{L51p1704} as observed in other critical phenomena. 
Using the reduced temperature $t$ = ($T$ $-$ $T_{\rm sg}$)/$T_{\rm sg}$ 
and a scaling function $F_\pm(z)$, the SG order parameter is defined as 
\begin{equation}
 q(T,H) = |t|^\beta F_\pm (H^2/|t|^\phi),
 \label{scaling}
\end{equation}
where $\beta$ and $\gamma = \phi - \beta$ 
are the critical exponents characterizing the SG state.\cite{L51p1704} 
The behavior of the scaling function $F_\pm(z)$ is well known 
in the following three limits: 
(i) $F_+(z \to 0)$ = 0 ($t>0$), (ii) $F_-(z \to 0)$ = const ($t<0$), 
and (iii) $F_{\pm}(z \to \infty)$ = $z^{\beta /\phi }$. 
Because a relation $q \sim t^\beta$ is immediately 
found from limit (ii), the exponent $\beta $ can be estimated from 
the slope in a log-log plot of $q$ (below $T_{\rm sg}$) versus $t$ in 
the limit of zero field. Such an analysis is performed in 
Fig.~\ref{scales}(a), and an estimate of $\beta $ = 0.7(1) is obtained. 
Similarly, using the limit (iii), the other exponent, 
$\phi$, can be determined from the dependence of $q$ on $H^2$ 
at $T_{\rm sg}$ from the relation $q(T_{\rm sg},H)$ = $(H^2)^{\beta/\phi}$. 
The slope in Fig.~\ref{scales}(b) then gives $\beta/\phi$ = 0.163(5) 
or $\phi$ = 4.3(8). 

Now that we have a good estimate of the critical exponents, we can
directly test the scaling relation described by Eq.~(\ref{scaling}). 
As shown in Fig.~\ref{scales}(c), we obtain excellent scaling. 
Through the process of getting the best scaling result, 
shown in Fig.~\ref{scales}(c), the critical exponents have 
been optimized further: $\beta = 0.78(5)$ and $\gamma = 4.1(5)$. 

\vspace*{0.5cm} 

The same SG features as for $x$ = 0.05 were observed 
for $x$ = 0.06 and 0.07. 
The critical exponents obtained in this manner do not depend on $x$, 
although $T_{\rm sg}$ decreases with $x$. 
We note that, in addition to the comparable values of $T_{\rm sg}$, 
the critical exponents are found to be almost identical to 
those of Sr-LCO,\cite{L75p2204,B62p3547} suggesting the existence 
of the same SG state in both systems. 
While the observed exponents are consistent with those of canonical SG 
materials,\cite{L75p2204} we note that 
recent results for untwinned Sr-LCO crystals reveal unconventional 
anisotropic behavior.\cite{Lavrov} 

\begin{figure}[b]
 \vspace*{0.8cm}
 \centerline{\epsfxsize=2.75in\epsfbox{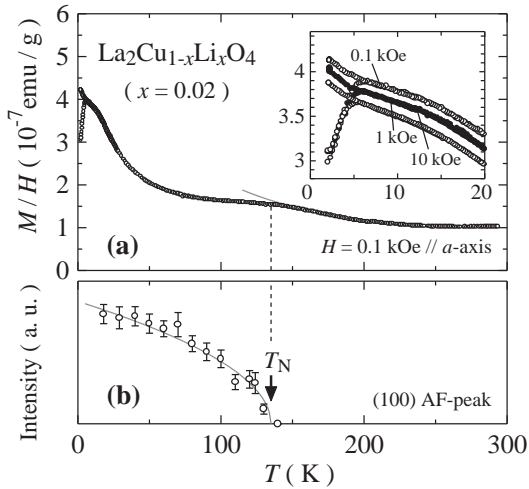}}
 \vspace*{0.3cm}
 \caption{(a) $\chi(T)$ of La$_2$Cu$_{0.98}$Li$_{0.02}$O$_4$ 
 as a function of temperature under $H$ = 0.1 kOe. 
 The inset magnifies the low-temperature behavior at various fields. 
 (b) $T$-dependence of the integrated intensity 
 of the (1 0 0) antiferromagnetic neutron peak (orthorhombic notation). 
 The power-law fit (solid curve) gives $T_N$ = 135 K, 
 which corresponds well to the cusp in the susceptibility.}
 \label{mag02}
\end{figure}

For $x <$ 0.03, the magnetic susceptibility differs from that 
for $x >$ 0.03. Data for $x$ = 0.02 are shown in 
Fig.~\ref{mag02}(a). Two magnetic anomalies are present. 
One is a high-temperature cusp 
associated with the AF transition 
which has already been observed in powder samples.\cite{B54p12014} 
As shown in Fig.~\ref{mag02}(b), neutron data complement the SQUID 
observations, providing unambiguous evidence of a well-defined N\'eel 
transition around this high-temperature bump. We have confirmed the
development of the antiferromagnetic peak at the (1,0,0) position
(orthorhombic notation) just below the susceptibility anomaly, and 
determined the N\'eel temperature to be $T_{\rm N}$ = 135 K.
The second anomaly, recognized well below $T_{\rm N}$, appears 
somewhat analogous to the SG behavior 
found for $x >$ 0.03 in that we find hysteresis below the anomaly 
and the onset temperature of magnetic irreversibility is 
comparable to $T_{\rm sg}$ for $x >$ 0.03. 
Neutron characterization furthermore revealed a sharp structural
transition from the tetragonal to orthorhombic phase at 490 K, 
consistent with a doping level of $x$ = 0.02.\cite{B54p12014} 
This indicates that our samples have a high degree of chemical homogeneity, 
and that the observed magnetic anomaly is intrinsic to the $x$ = 0.02 phase. 
Local-probe $^{139}$La-NQR and $\mu$SR experiments 
have reported a corresponding spin-freezing 
within the AF state in Sr-LCO (Refs. 3 and 9) 
and Li-LCO.\cite{L81p2791} 
This has been ascribed to the continuous freezing of the spins of 
the doped holes on the antiferromagnetic background,\cite{L71p2323} 
with independent ordering of Cu$^{2+}$ spins and doped holes,\cite{L80p3843} 
or due to a collective hole behavior.\cite{L81p2791} 
However, to the best of our knowledge, 
this spin freezing has not been reported thus far from 
bulk magnetometry for the cuprates. 
Because the SG behavior is very sensitive to the dopant 
concentration ($x$-dependence will be discussed later) and 
to the field direction, the success of the present observations 
depended greatly on high-quality single crystals as well as 
extended measurements in the low-temperature region. 

Although the second anomaly appears to be a SG transition, further 
characterization in terms of critical scaling analysis is not possible 
due to the ambiguity in the definition of the SG order parameter. 
Nevertheless, it would be worthwhile to determine a characteristic 
freezing temperature, $T_f$. 
We define $T_f$ as the lower-temperature ``shoulder" 
of the zero-field-cooled data at low $H$, 
which corresponds to the onset of magnetic hysteresis for $x$ = 0.02, 
as seen in the inset of Fig.~\ref{mag02}(a). 
We note that, although not shown here, a change in the imaginary part 
of the ac susceptibility was also observed at this temperature, 
which can be associated with the drastic change of the spin response 
due to the onset of the spin freezing. 

\begin{figure}[t]
 \centerline{\epsfxsize=2.2in\epsfbox{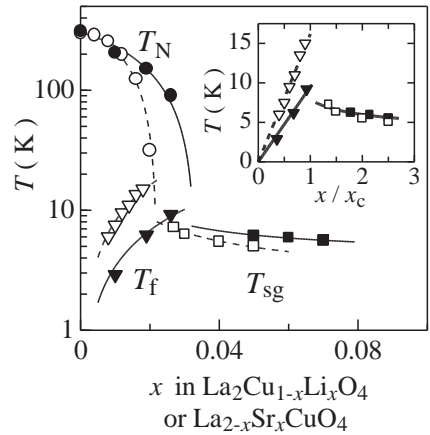}}
 \vspace*{0.3cm}
  \caption{Magnetic phase diagram for La$_2$Cu$_{1-x}$Li$_x$O$_4$ 
 (filled symbols) and La$_{2-x}$Sr$_x$CuO$_4$ (open symbols; from 
 NQR $(x<0.02)$\protect\cite{L71p2323} and 
 magnetometry $(x>0.02)$\protect\cite{B62p3547}): 
 circles for the N\'eel temperature $T_N$, 
 triangles for the onset of the SG-like anomaly $T_{\rm f}$, 
 and squares for the SG temperature $T_{\rm sg}$.}
 Inset: scaled phase diagram.
\label{phase}
\end{figure}
\vspace*{0.5cm} 

Our results yield the phase diagram summarized in Fig.~\ref{phase}. 
NQR ($x<0.02$) (Ref. 3) and magnetometry ($x>0.02$) (Ref. 7) 
data from Sr-LCO single crystals are included for comparison. 
Despite the different nature of the dopants, the phase boundaries 
for these two compounds are almost identical. 
Figure~\ref{phase} demonstrates that $T_{\rm f} \sim x$ as in Sr-LCO, 
opposite to the trend with doping for either $T_N$ or $T_{\rm sg}$. 
We find $T_{\rm f}$ = (339 (14) K)$\,\times \,x$, as compared to 
$T_{\rm f} \approx$ (815 K)$\,\times \,x$ 
for Sr-LCO from $^{139}$La NQR.\cite{L71p2323} 

The dopant concentrations at which N\'eel order is lost 
in the two materials differ by a factor of $\sim $\,2/3. 
Qualitatively, this shift of the phase boundaries 
can be understood to result from reduced 
magnetic frustration in the Li-doped system.\cite{Korenblit} 
As shown in the inset of Fig.~\ref{phase}, when normalized by that 
factor, $T_{\rm sg}$ for Li-LCO ($x > 0.03$) 
$quantitatively$ agrees with the SG temperature for Sr-LCO. 
Since NQR and magnetometry probe very different time scales, 
systematic Sr-LCO magnetometry data for $x<0.02$ would be very 
desirable for a proper quantitative comparison at low doping. 

Recent neutron scattering results reveal direct evidence for 
electronic phase separation in Sr-LCO ($x < 0.02$) into regions with 
hole concentrations $\approx 0$ and $\approx 0.02$.\cite {Matsuda} 
The latter phase exhibits diagonal stripes.\cite{B62p3547} 
Spin freezing occurs below the doping-independent 
phase separation temperature $T_{ps} \approx 30$ K determined from 
neutron scattering,\cite{Matsuda} and consistent 
with previous NQR results.\cite{L71p2323} 
For Li-LCO, on the other hand, commensurate AF correlations have been 
both predicted\cite{Buhler} and observed,\cite{Bao} consistent with 
the expected stronger pinning potential of the in-plane dopant Li$^+$. 
Consequently, the cluster model, originally 
proposed for Sr-LCO, but which predicts commensurate AF correlations, 
might more accurately describe the physics of 
the Li-doped variant.\cite{Gooding} 
This model predicts $k_{\rm B}T_{\rm f} \sim J_{\rm eff}x$, 
where $J_{\rm eff}$ is the effective in-plane exchange coupling 
constant.\cite{Gooding} 
While this linear doping dependence is indeed consistent with our 
observations, the same behavior is found in Sr-LCO.\cite{L71p2323} 
On the other hand, it has been speculated that $T_{\rm f}$ in Sr-LCO 
might depend linearly on the volume 
fraction of the SG phase.\cite{Matsuda} 
The close analogy between the NQR (Refs. 3 and 22) 
and bulk magnetic properties of the two materials suggests 
that Li-LCO might phase separate as well. 

At higher hole concentrations, the behavior of the carriers differs 
significantly between Li-LCO and Sr-LCO. The latter 
material shows an insulator to metal transition and HTS, 
while the former remains insulating. 
What is particularly interesting from our present observations is that, 
in spite of the great difference in charge dynamics for $x >$ 0.03, 
the spin degrees of freedom measured via magnetometry are remarkably 
similar in both compounds. 
Both materials show a spin glass transition, 
with comparable transition temperatures at the same doping value 
and with the same critical exponents. 
The values of $T_{\rm sg}(x)$ for the more disordered material 
Li-LCO lie above those of Sr-LCO. 
The relative shift of the SG phase boundaries likely 
results from an effective decrease of magnetic frustration due to 
the presence of nonmagnetic Li$^+$ in Li-LCO.\cite{Korenblit} 
In this context, it is worth noting that the magnetic phase boundaries 
in the double-layer cuprates\cite{L80p3843,PC341p843} lie 
above those of both Li-LCO and Sr-LCO. 
While differences in the strength and type of disorder might 
play a subtle role in setting the temperature scales 
for N\'eel and SG order, we believe that the predominant effect 
is the difference in the effective three-dimensional AF coupling 
between single- and double-layer materials. 
Specifically, while the interplanar AF coupling is nearly frustrated 
in doped LCO, this is not the case for the double-layer materials. 
On the other hand, the relative insensitivity of 
the low-temperature magnetic phase boundaries of the structurally 
identical materials Li-LCO and Sr-LCO to the type and strength of 
the quenched disorder is consistent with the notion that the glassiness 
in these doped Mott insulators is primarily self-generated.\cite{L85p836} 

\section{CONCLUSION} 
The bulk magnetic properties of Li-LCO single crystals 
are found to be richer than previously reported from powder samples. 
Our results demonstrate the feasibility of extracting low-temperature 
freezing temperatures in the N\'eel state from magnetometry, and call 
for similar measurements in other cuprates 
for a proper quantitative comparison. 
We find that $T_{\rm f} \sim x$, 
which also has been reported for Sr-LCO. 
Since the spin freezing in the latter material occurs 
in a phase separated state, the close similarity between the doping 
dependence of the freezing temperature 
suggests that lightly doped Li-LCO might phase separate as well, 
despite the stronger pinning potential of the in-plane dopant Li$^+$. 
Outside of the N\'eel phase, the spin glass phase boundaries differ 
by a mere scale factor for the effective doping level. 
The experimental results for Li-LCO obtained here extend the universality 
of the bulk magnetic phase diagram to hole-doped 
but non-superconducting cuprates. 

\section*{ACKNOWLEDGMENTS}
We thank D. Fisher, H. Eisaki, K. Kishio, R.S. Markiewicz, 
Ch. Niedermayer, C. Panagopoulos, and H. Takagi 
for fruitful discussions. 
P.K.M., O.P.V., and M.G. would like to thank the staff 
of the NIST Center for Neutron Research for their warm hospitality. 
The work at Stanford was supported by the U.S. DOE 
under Contract Nos. DE-FG03-99ER45773 and DE-AC03-76SF00515, 
by NSF CAREER Award No. DMR-9985067, and by the A.P. Sloan Foundation. 
T.S. was partially supported by JSPS. 


\end{document}